\documentclass[a4paper,12pt]{article}

\usepackage[cp1250]{inputenc} %Czech and Slovak symbols
\usepackage{authblk}%Formating of authors' information on title page
\usepackage{graphicx}
\usepackage{amsfonts}
\usepackage{amsmath}
\usepackage{amsthm}
\usepackage{amssymb}
\usepackage{booktabs}
\usepackage{a4wide}
\usepackage[T1]{fontenc}

\begin{document}

\title{Fractal Markets Hypothesis and the Global Financial Crisis: Wavelet Power Evidence\footnote{Published as Kristoufek, L. (2013): Fractal Markets Hypothesis and the Global Financial
Crisis: Wavelet Power Evidence. \textit{Scientific Reports} \textbf{3}, 2857; DOI:10.1038/srep02857. Paper is available at http://www.nature.com/srep/2013/131004/srep02857/full/srep02857.html}}

\author[a,b]{Ladislav Kristoufek}
\affil[a]{Institute of Economic Studies, Faculty of Social Sciences, Charles University in Prague, Opletalova 26, 110 00, Prague, Czech Republic, EU}
\affil[b]{Institute of Information Theory and Automation, Academy of Sciences of the Czech Republic, Pod Vodarenskou vezi 4, 182 08, Prague, Czech Republic, EU, tel.: +420266052243, Correspondence to kristouf@utia.cas.cz}
\date{\today}
\maketitle

\begin{abstract}
\footnotesize
We analyze whether the prediction of the fractal markets hypothesis about a dominance of specific investment horizons during turbulent times holds. To do so, we utilize the continuous wavelet transform analysis and obtained wavelet power spectra which give the crucial information about the variance distribution across scales and its evolution in time. We show that the most turbulent times of the Global Financial Crisis can be very well characterized by the dominance of short investment horizons which is in hand with the assertions of the fractal markets hypothesis.
\end{abstract}

\normalsize

Critical events and turbulences on the financial markets have always attracted attention of financial researchers as these are the biggest challenges for prevailing theories and potentially good starting points for new paradigms. The efficient markets hypothesis \cite{Fama1970} has been a cornerstone of mainstream financial economics for decades even though it provides no testable predictions about extreme events and even considers them highly improbable (or even non-existent) \cite{LeRoy1976,Stanley2003}. In this paper, we try to resurrect the practically forgotten fractal markets hypothesis \cite{Peters1994} and test whether its predictions about dominance of specific investment horizons during the turbulent times hold in practice. To do so, we assume that an increased volatility is a reflection of an increased trading activity \cite{Karpoff1987,Jain1988} and utilize the continuous wavelet transform analysis to get the wavelet power spectra, which can be interpreted as a scale- and time-characteristic variance. Wavelets have been widely used across various disciplines -- cardiology \cite{Thurner1998,Papasimakis2010}, genetics \cite{Arneodo1995,Machado2009}, geophysics \cite{Grinsted2004}, neurology \cite{Hulata2000}, traffic modeling \cite{Huang2003}, economics \cite{Aguiar-Conraria2008} and others. Using wavelets methodology, we are able to comment on evolution of trading activity in time but importantly also across scales and we show that predictions of the fractal markets hypothesis are indeed observed during the latest turbulences on the financial markets.

The current Global Financial Crisis (2007 onwards) has once again showed that an ongoing growth of financial asset prices is not limitless. Commenced in the USA, the uncertainty about the future economic prospects has spread globally. During the most turbulent times, the major stock indices decreased considerably and lost usually around 50\% of their 2007 peak values during the 1.5 year long decline. According to the mainstream efficient markets hypothesis (EMH), such a crisis should have never happened because investors are assumed to price the assets rationally and with respect to all available information \cite{Fama1970,Malkiel2003}. Moreover, investors are assumed to be homogeneous so that the arriving information has the same effect on all investors. However, it is well known that market participants are greatly heterogenous with investment horizons ranging from a fraction of seconds and minutes (market makers, noise-traders) up to several years (pension funds). Investors with short investment horizons focus on technical information and crowd behavior of other market participants. On the other hand, investors with long investment horizons base their decisions on fundamental information and care little about the crowd behavior. In turn, the financial markets become highly complex systems, which are rather hard to be analyzed with standard linear methods.

The fractal markets hypothesis (FMH), on contrary, is based on empirically observed characteristics of the financial markets and thus considers the market to consist of heterogenous agents who react to the inflowing information with respect to their investment horizon \cite{Peters1994,Rachev1999,Weron2000}. What is considered a negative information and thus a selling signal for an investor with short horizon might be a buying opportunity for an investor with long horizon, and vice versa. If a sufficient number of buyers and sellers trade and are efficiently cleared in the market mechanism, a smooth functioning of this market is guaranteed. This brings us to the crucial notion of FMH -- liquidity. If investment horizons are uniformly represented in the market (i.e. there is a sufficient number of investors at a wide range of investment horizons), then supply and demand for assets are met, market works efficiently and remains stable. However, if an investment horizon (or a group of horizons) becomes dominant, buying and selling orders are not efficiently cleared and extreme events are likely to occur \cite{Farmer2004,Weber2006,Mike2008,Roch2011}. Therefore, FMH directly predicts that critical events are connected to dominating investment horizons. It is then straightforward that short investment horizons should dominate during financial turmoils and that is for two possible reasons -- (i) the long-term investors panic and start selling as a reaction to the negative news and crowd behavior, and in turn activity at short horizons increases compared to the long ones, and/or (ii) the long-term investors stay out of the market until the situation calms making the short horizons dominant. To uncover whether this prediction of FMH is correct for the Global Financial Crisis (GFC) of the late 2000s, we apply the continuous wavelet transform procedure on developed and liquid world stock indices -- the USA is represented by NASDAQ, the UK by FTSE, continental Europe by French CAC and German DAX, and Asia by HSI of Hong Kong and Japanese NIKKEI -- between 3.1.2000 and 31.5.2013.

The basic idea behind application of wavelets on the financial processes is their ability to analyze the underlying process both in the time and frequency domain. In a similar manner as the standard Fourier analysis, the series is decomposed into frequencies and a scale-specific power is obtained. However, the wavelets add a significant improvement over the Fourier analysis as they provide a way to study how this scale-specific power evolves in time (for more technical details, see the Methods section). In terms of financial economics, the wavelet power can be understood as a scale-specific variance. Compared to the standardly used techniques such as various versions of (generalized) autoregressive conditional heteroskedasticity models \cite{Engle1982,Bollerslev1986}, we obtain information about time evolution of variance as well as its distribution across frequencies (scales). According to the FMH arguments, we should observe increased power at low scales (high frequencies) during the critical periods. Moreover, we might observe a changing structure of variance across frequencies before the turbulences due to the changing structure of investors' activity as reported in our previous study \cite{Kristoufek2012}.

\section*{Results}

\begin{figure}[htbp]
\center
\begin{tabular}{c}
\includegraphics[width=5.5in]{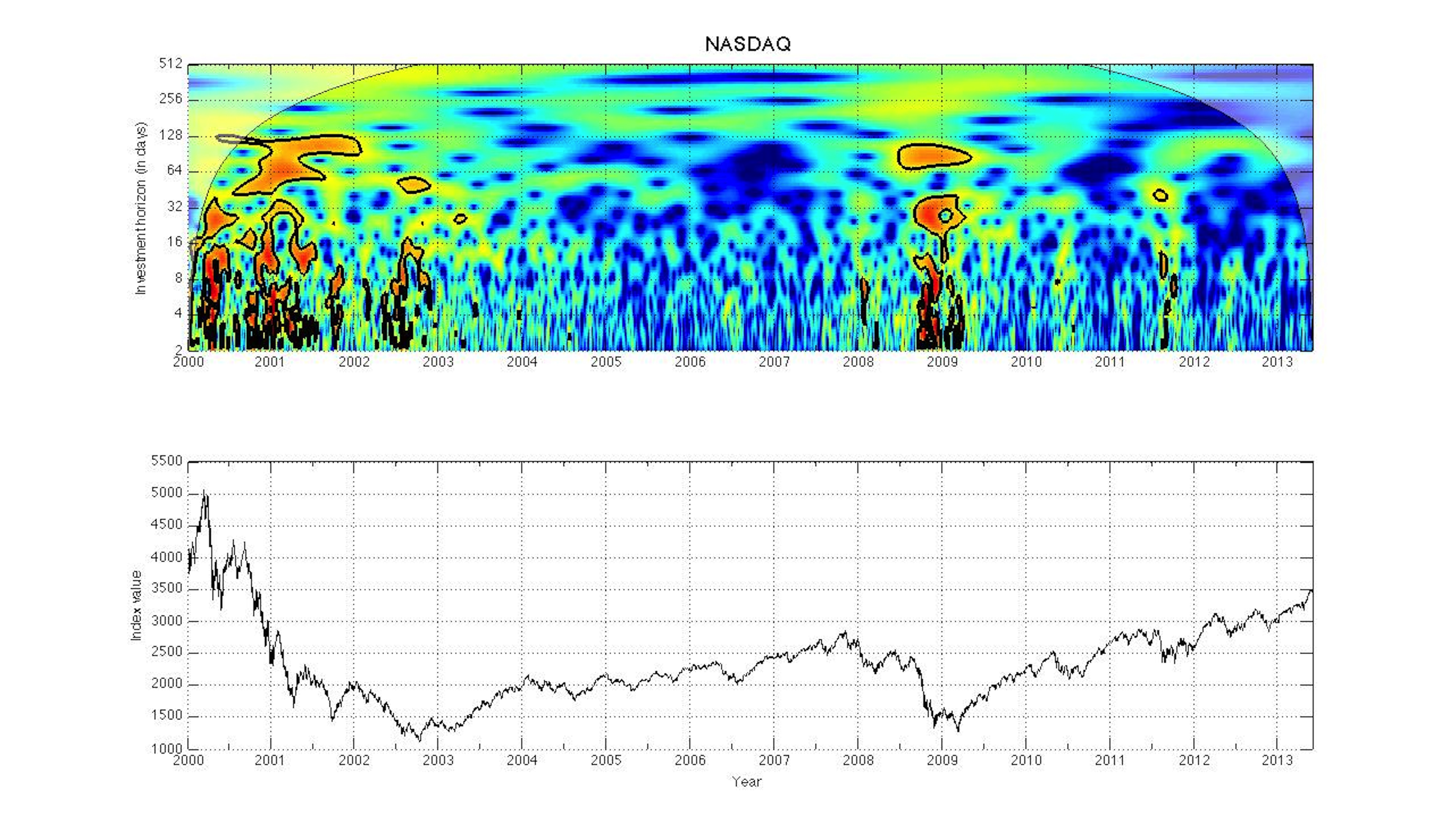}\\
\end{tabular}
\caption{\footnotesize\textit{Wavelet power spectrum for NASDAQ.} Significant wavelet powers against the null hypothesis of a red noise are marked off by a bold black line. Cone of influence separates the spectrum into two -- a top (pale) part where the inference is less reliable and a bottom part (colorful) where the results are reliable. The hotter the color (as shown on the scale at the right-hand side of the charts), the higher the power (variance) at the specific scale ($y$-axis) and time ($x$-axis). Statistically significant regions are bordered by a thick black curve. As the data resolution is daily, the scales are in day units. Wavelet power is shown for the whole analyzed period (upper chart) together with evolution of the index value (bottom chart). Dominance of high frequencies is evident for the most turbulent times of the GFC but also for other critical trend changes and corrections so that the results are in hand with fractal markets hypothesis predictions.}\label{fig1}
\end{figure}

\begin{figure}[htbp]
\center
\begin{tabular}{c}
\includegraphics[width=5.5in]{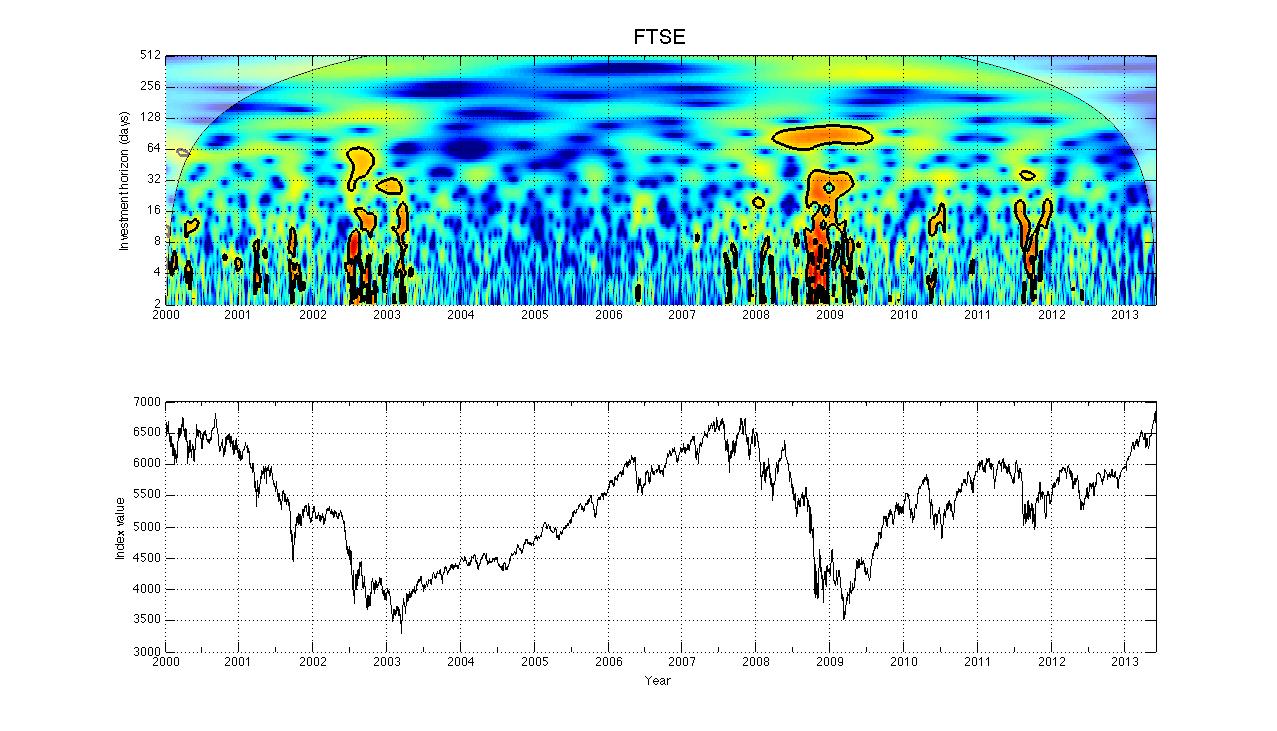}\\
\end{tabular}
\caption{\footnotesize\textit{Wavelet power spectrum for FTSE.} Labeling and qualitative results hold from the NASDAQ case.}\label{fig2}
\end{figure}

\begin{figure}[htbp]
\center
\begin{tabular}{c}
\includegraphics[width=5.5in]{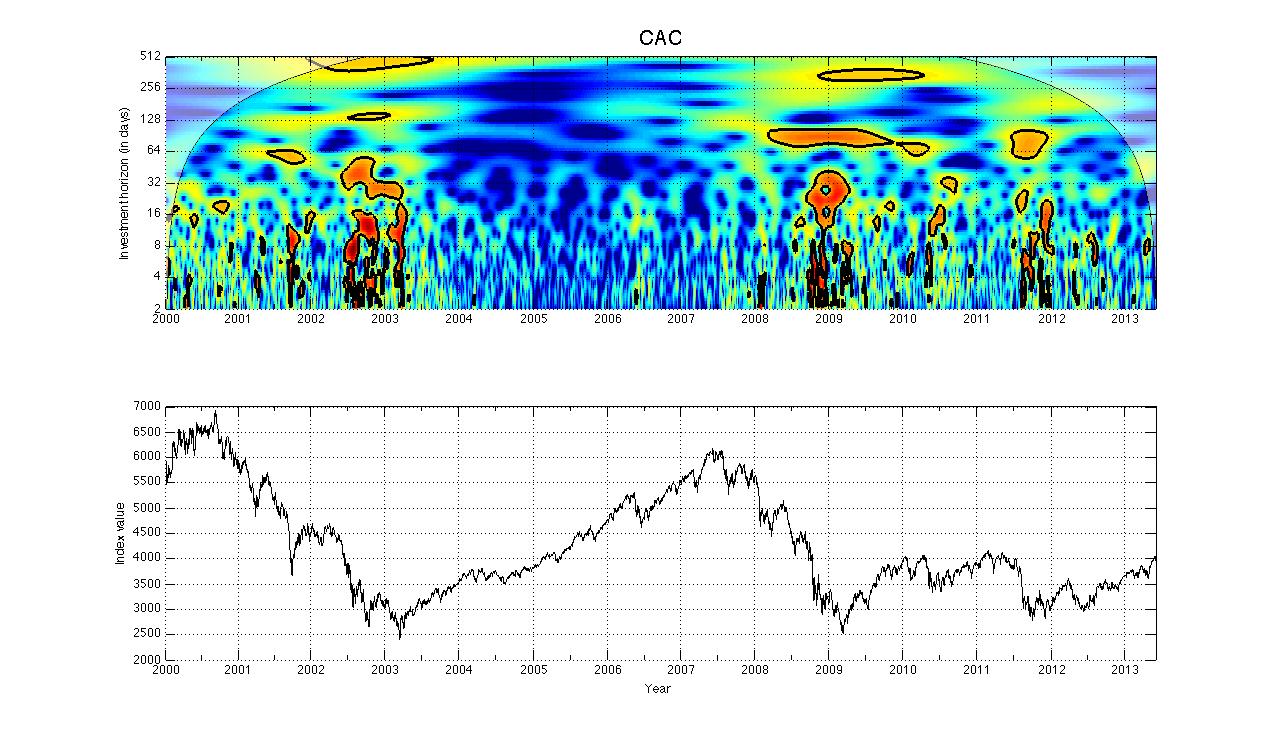}\\
\end{tabular}
\caption{\footnotesize\textit{Wavelet power spectrum for CAC.} Labeling and qualitative results hold from the NASDAQ case.}\label{fig3}
\end{figure}

\begin{figure}[htbp]
\center
\begin{tabular}{c}
\includegraphics[width=5.5in]{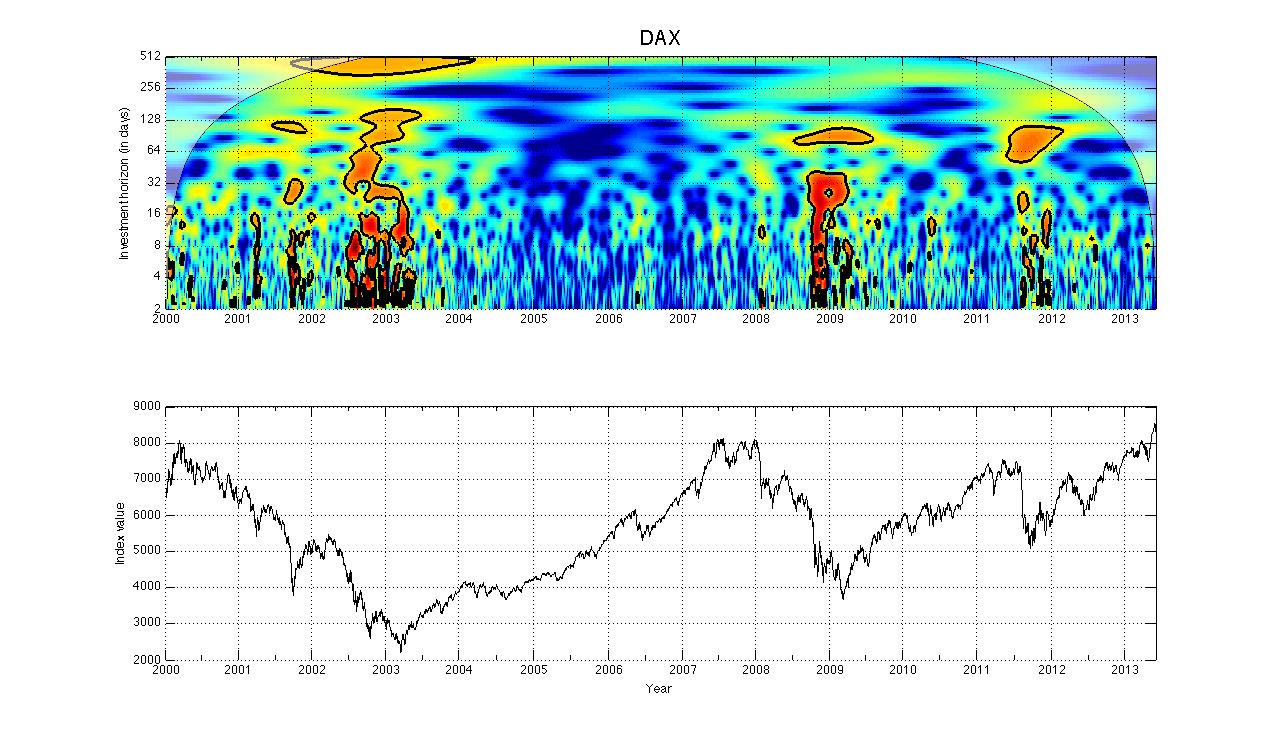}\\
\end{tabular}
\caption{\footnotesize\textit{Wavelet power spectrum for DAX.} Labeling and qualitative results hold from the NASDAQ case.}\label{fig4}
\end{figure}

\begin{figure}[htbp]
\center
\begin{tabular}{c}
\includegraphics[width=5.5in]{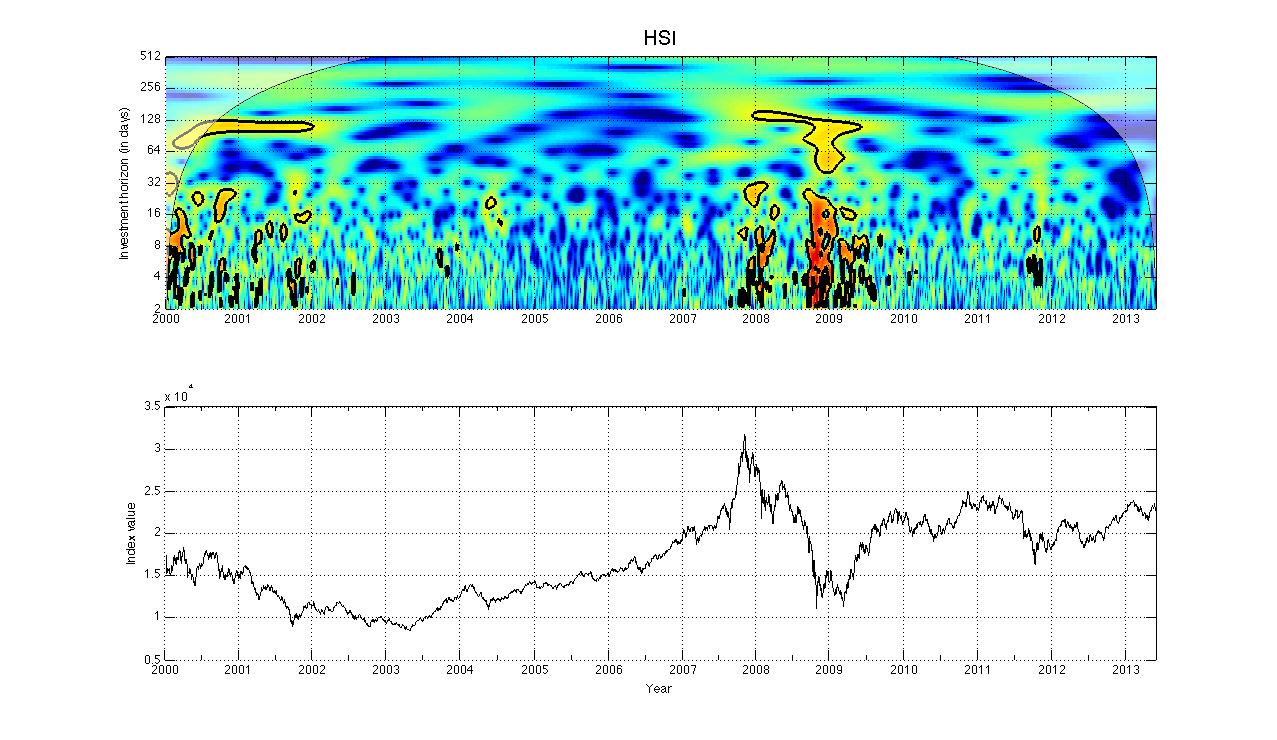}\\
\end{tabular}
\caption{\footnotesize\textit{Wavelet power spectrum for HSI.} Labeling and qualitative results hold from the NASDAQ case.}\label{fig5}
\end{figure}

\begin{figure}[htbp]
\center
\begin{tabular}{c}
\includegraphics[width=5.5in]{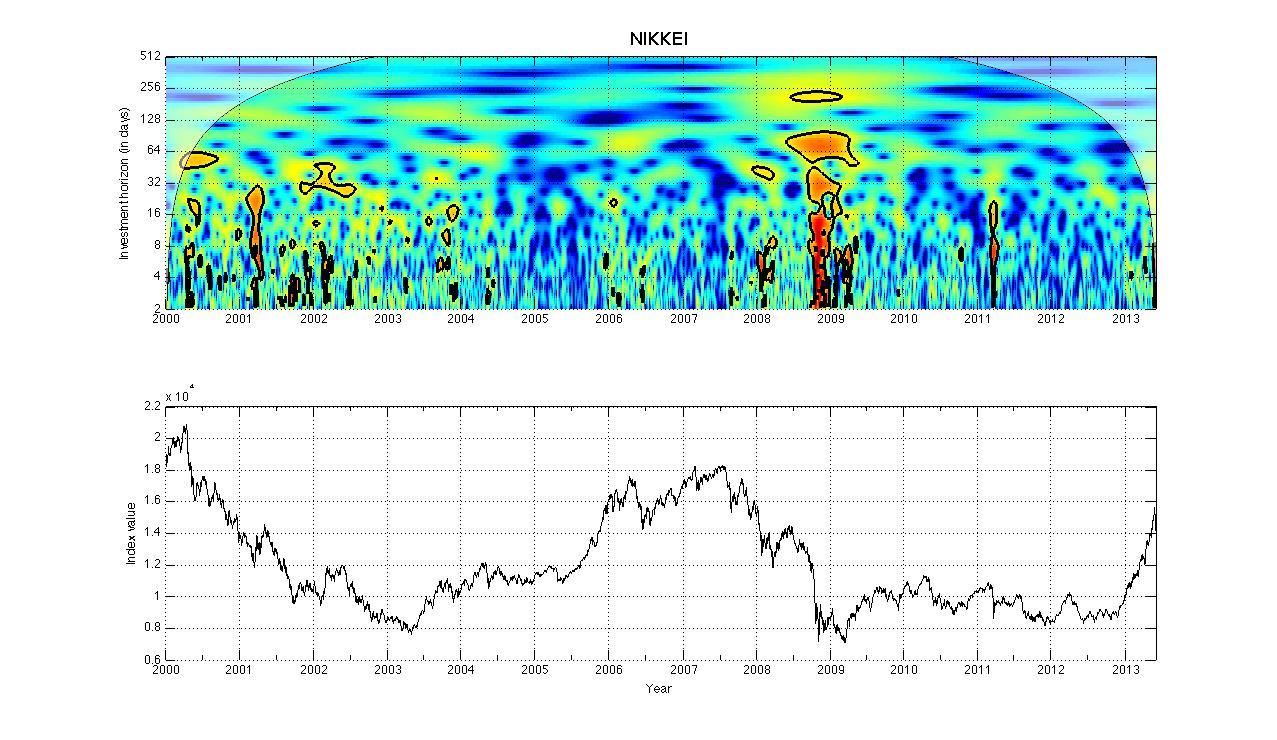}\\
\end{tabular}
\caption{\footnotesize\textit{Wavelet power spectrum for NIKKEI.} Labeling and qualitative results hold from the NASDAQ case.}\label{fig6}
\end{figure}

To show whether the predictions of the fractal markets hypothesis hold, we analyze six developed (liquid) world indices. The USA is represented by NASDAQ Composite Index (NASDAQ). The UK is covered by FTSE 100 Index (FTSE). Continental Europe is represented by stock indices of two most developed Eurozone countries -- German DAX (DAX) and French CAC 40 (CAC). Asia is represented by two stock indices -- Hang Seng Index (HSI) of Hong Kong and Japanese NIKKEI 225 (NIKKEI). The analyzed period ranges between 3.1.2000 to 30.5.2013, which attributes around 3400 observations for each index. Evolution of index values is illustrated in Figs. \ref{fig1}-\ref{fig6}. We can see that the indices have evolved similarly but there are several distinct features for each market. 

NASDAQ as primarily a technological index was most severely hit by the DotCom bubble at the break of the millennium and it plummeted from its highs around 5000 points down to 1000 points at the end of 2002 when a slow but steady increasing trend began and persisted up to the point of the current Global Financial Crisis. Regardless the after-DotCom growth, the heights reached before the current crisis were only around a half of the 2000 peak. FTSE index underwent a similar path loosing around 50\% during the period after the DotCom bubble. Compared to NASDAQ, the decrease was more gradual. In a similar manner, the DotCom deflation of the index was followed by a continuous growth which was much more profound compared to NASDAQ as the 2000 heights were reached in the half of 2007. After the Lehman Brothers collapse, the index plunged to the levels experienced after the DotCom bubble. However, the index has recovered quite rapidly and as of half of 2013, the peaks of 2000 and 2007 have been reached again. In comparison to the NASDAQ index, FTSE experienced an additional correction in the second half of 2011. This correction can be seen for all European indices and it is connected to the debt crisis of the EU (the Greek crisis, the PIIGS -- Portugal, Italy, Ireland, Greece, Spain -- crisis, and the Eurozone crisis to mention some other names). This dip in the index value is evident also for CAC and DAX but not so much for the non-European NASDAQ, HSI and NIKKEI. CAC and DAX indices can be described in the same words as the FTSE index which shows the interconnection between the European markets. HSI index shows a very different picture. Even though the global downturn of the beginning of the millennium is reflected here as well, the surge of the index after 2003 up to late 2007 is much more profound. Nonetheless, the Global Financial Crisis breakdown is evident here in the same proportions as for the other indices. Even until the half of 2013, the index has not recovered and the market has stabilized at around 65\% of its 2007 peak value. NIKKEI follows the dynamics of the European indices in the beginning of 2000s and of the NASDAQ index during the GFC. The market was not able to recover from the GFC downturn for practically four years and only recently, it has started an explosive growth since the end of 2012 most likely connected to the changing direction of the economic policies in Japan.

Figs. \ref{fig1}-\ref{fig6} depict the wavelet power spectra for the logarithmic returns of the analyzed indices. In the charts, the significance of the wavelet powers is tested against the null hypothesis of a red noise (AR(1) process) \cite{Allen1996,Grinsted2004}, which is considered sufficient for the financial series -- areas of the significant power are bordered by a bold black curve. For finite series as the ones analyzed, we arrive at the limitation of wavelets at high scales where the transforms cannot be efficiently calculated. The so-called cone of influence marked with paler colors separates the time-frequency space into two -- the lower one where the inference is reliable and the upper one where it is not. For the interpretation of the powers, the hotter the color the higher the power (variance) for the particular scale and time.

For all the analyzed indices, we observe at least two common features. Firstly, most of the time, there is no scale-characteristic power which can be distinguished from a pure noise so that no investment horizon dominates. This is true mainly for the period between 2003 and the first half of 2008. Note that this period is connected to a very stable growth after the long-lasting decline following the DotCom bubble. The period is thus connected to a uniformly distributed activity at separate investment horizons (or at least, there is no dominating horizon), which is well in hand with FMH. The whole period between 2003 and 2008 is characterized by low total variance at all frequencies and the colors of the power spectra remain cold. Secondly, the turning point between the post-DotCom decreases and the upcoming upward trends is connected with an increased power at high frequencies. This is true for the European and the US indices, the Asian indices do not experience such behavior. Note that for all the indices, the frequent dips in the market are connected with a short-term increased power at the highest frequencies, which shows that these short-lived corrections of the markets are mainly dominated by the short-term investors. Note also that the wavelet power of NASDAQ displays a rather erratic behavior during the DotCom bubble which only highlights the fact that NASDAQ has been the market which was most severely hit by the DotCom bubble bursting.

Turning now to the current Global Financial Crisis, recall that the biggest movements occurred after the bankruptcy of Lehman Brothers on 15.9.2008. In Figs. \ref{fig1}-\ref{fig6}, we can see that mainly the period between September and December 2008 is strongly dominated by an increased energy at the highest frequencies. Even though the variance increases at more scales, the dominance of the very low scales is apparent (very hot colors even bordering with black for October and November 2008) and is again in hand with the predictions of FMH -- the turbulent times are connected with the dominance of specific investment horizons so that the efficient market clearing is not possible. The situation calmed around May 2009 where the total variance decreased markedly.

Apart from the most severe periods of the GFC, we also observe a changing structure of the active participants at the European stock markets during the second half of 2012. Note that during this period, the Eurozone and the EU in general underwent probably the most severe financial and economic turmoil of its existence. The situation is known by various names as the symptoms were many -- the EU debt crisis, the PIIGS (for Portugal, Ireland, Italy, Greece and Spain as the most troublesome countries in that times) crisis, the Greek crisis and several more. The fact that this crisis was geographically specific is well illustrated in its effects being seen only for CAC, DAX and FTSE. Again the dominance of the shortest investment horizons is evident. 

\section*{Discussion}

In summary, we have analyzed whether the prediction of the fractal markets hypothesis about the dominance of specific investment horizons during the turbulent times holds. To do so, we have utilized the continuous wavelet transform analysis and the wavelet power spectra which give crucial information about the variance distribution across scales and its evolution in time. We have showed that the most turbulent times of the current Global Financial Crisis can be very well characterized by the dominance of short investment horizons which is well in hand with the fractal markets hypothesis. Misbalance between short and long investment horizons thus created a tension between supply and demand, leading to decreased liquidity which has been repeatedly shown to lead to occurrence of extreme events \cite{Farmer2004,Weber2006,Mike2008,Roch2011}. Fractal markets hypothesis is thus able to describe events of the Global Financial Crisis in a more satisfying way than the mainstream efficient markets hypothesis.

%It is thus likely that the crisis has been connected with a misbalance between supply and demand for traded securities leading to extreme movements and uncertainty. However, to test such an assertion empirically, additional control variables directly measuring discrepancies in supply and demand, such as bid-ask prices and spreads, would be needed. The results thus serve as a starting point for further research of liquidity drains at various investment horizons during the critical events, namely the Global Financial Crisis.

\section*{Methods}

\subsection*{Data}

Daily logarithmic returns $r_t=\log(C_t)-\log(O_t)$, where $C_t$ and $O_t$ are closing and opening prices, respectively, are analyzed between 3.1.2000 and 31.5.2013. The US stock markets are represented by NASDAQ Composite Index (ticker symbol $IXIC$). The European stock markets are covered by CAC 40 ($FCHI$), DAX ($GDAXI$) and FTSE 100 ($FTSE$) and the Asian ones by Hang Seng Index ($HSI$) and NIKKEI 225 ($N225$). All series were obtained from http://finance.yahoo.com.

\subsection*{Wavelet}

A wavelet $\psi_{u,s}(t)$ is a real-valued square integrable function defined as

\begin{equation}
\psi_{u,s}(t)=\frac{\psi\left(\frac{t-u}{s}\right)}{\sqrt{s}}
\end{equation}
with scale $s$ and location $u$ at time $t$. Any time series can be reconstructed back from its wavelet transform if the admissibility condition
\begin{equation}
C_{\Psi}=\int_0^{+\infty}{\frac{|\Psi(f)|^2}{f}df}<+\infty
\end{equation}
holds, where $\Psi(f)$ is the Fourier transform of a wavelet. Wavelet has a zero mean so that $\int_{-\infty}^{+\infty}{\psi(t)}dt=0$ and is usually normalized so that $\int_{-\infty}^{+\infty}{\psi^2(t)}dt=1$. To obtain the continuous wavelet transform $W_x(u,s)$, a wavelet $\psi(.)$ is projected onto the examined series $x(t)$ so that
\begin{equation}
W_x(u,s)=\int_{-\infty}^{+\infty}{\frac{x(t)\psi^{\ast}\left(\frac{t-u}{s}\right)dt}{\sqrt{s}}}
\end{equation}
where $\psi^{\ast}(.)$ is a complex conjugate of $\psi(.)$. Importantly, the continuous wavelet transform decomposes the series into frequencies and can then reconstruct the original series so that there is no information loss, and energy of the examined series is maintained as well, i.e.
\begin{equation}
x(t)=\frac{\int_0^{+\infty}{\int_{-\infty}^{+\infty}{W_x(u,s)\psi_{u,s}(t)du}ds}}{s^2C_{\Psi}},
\end{equation}
\begin{equation}
||x||^2=\frac{\int_0^{+\infty}{\int_{-\infty}^{+\infty}{|W_x(u,s)|^2du}ds}}{s^2C_{\Psi}}
\end{equation}
where $|W_x(u,s)|^2$ is the wavelet power at scale $s>0$. 

There is a large number of specific wavelets \cite{Percival2000} out of which we choose the Morlet wavelet standardly used in the economic and financial applications \cite{Aguiar-Conraria2008,Rua2009,Vacha2012,Vacha2013}. The Morlet wavelet with a central frequency $\omega_0$ is defined as
\begin{equation}
\psi(t)=\frac{e^{i\omega_0t-t^2/2}}{\pi^{1/4}}
\end{equation}
and for $\omega_0=6$, it provides a good balance between the time and frequency localization \cite{Grinsted2004}.

\bibliographystyle{naturemag}
%\bibliography{FMH}

\begin{thebibliography}{10}
\expandafter\ifx\csname url\endcsname\relax
  \def\url#1{\texttt{#1}}\fi
\expandafter\ifx\csname urlprefix\endcsname\relax\def\urlprefix{URL }\fi
\providecommand{\bibinfo}[2]{#2}
\providecommand{\eprint}[2][]{\url{#2}}

\bibitem{Fama1970}
\bibinfo{author}{Fama, E.}
\newblock \bibinfo{title}{{Efficient Capital Markets: A Review of Theory and
  Empirical Work}}.
\newblock \emph{\bibinfo{journal}{J. Financ.}}
  \textbf{\bibinfo{volume}{25}}, \bibinfo{pages}{383--417}
  (\bibinfo{year}{1970}).

\bibitem{LeRoy1976}
\bibinfo{author}{LeRoy, S.}
\newblock \bibinfo{title}{Efficient capital markets: Comment}.
\newblock \emph{\bibinfo{journal}{J. Financ.}}
  \textbf{\bibinfo{volume}{31}}, \bibinfo{pages}{139--141}
  (\bibinfo{year}{1976}).

\bibitem{Stanley2003}
\bibinfo{author}{Stanley, H.~E.}
\newblock \bibinfo{title}{Statistical physics and economic fluctuations: do
  outliers exist?}
\newblock \emph{\bibinfo{journal}{Physica A}} \textbf{\bibinfo{volume}{318}},
  \bibinfo{pages}{279--292} (\bibinfo{year}{2003}).

\bibitem{Peters1994}
\bibinfo{author}{Peters, E.}
\newblock \emph{\bibinfo{title}{{Fractal Market Analysis -- Applying Chaos
  Theory to Investment and Analysis}}} (\bibinfo{publisher}{John Wiley \& Sons,
  Inc.}, \bibinfo{address}{New York}, \bibinfo{year}{1994}).

\bibitem{Karpoff1987}
\bibinfo{author}{Karpoff, J.}
\newblock \bibinfo{title}{The relation between price changes and trading
  volume: A survey}.
\newblock \emph{\bibinfo{journal}{J. Financ. Quant. Anal.}} \textbf{\bibinfo{volume}{22}}, \bibinfo{pages}{109--126}
  (\bibinfo{year}{1987}).

\bibitem{Jain1988}
\bibinfo{author}{Jain, P.} \& \bibinfo{author}{Joh, G.-H.}
\newblock \bibinfo{title}{The dependence between hourly proces and trading
  volume}.
\newblock \emph{\bibinfo{journal}{J. Financ. Quant. Anal.}} \textbf{\bibinfo{volume}{23}}, \bibinfo{pages}{269--283}
  (\bibinfo{year}{1988}).

\bibitem{Thurner1998}
\bibinfo{author}{Thurner, S.}, \bibinfo{author}{Feurstein, M.} \&
  \bibinfo{author}{Teich, M.}
\newblock \bibinfo{title}{Multiresolution wavelet analysis of heartbeat
  intervals discriminates healthy patiens from those with cardia pathology}.
\newblock \emph{\bibinfo{journal}{Phys. Rev. Lett.}}
  \textbf{\bibinfo{volume}{80}}, \bibinfo{pages}{1544--1547}
  (\bibinfo{year}{1998}).

\bibitem{Papasimakis2010}
\bibinfo{author}{Papasimakis, N.} \& \bibinfo{author}{Pallikari, F.}
\newblock \bibinfo{title}{Correlated and uncorrelated heart rate fluctuations
  during relaxing visualization}.
\newblock \emph{\bibinfo{journal}{EPL}} \textbf{\bibinfo{volume}{90}},
  \bibinfo{pages}{48003} (\bibinfo{year}{2010}).

\bibitem{Arneodo1995}
\bibinfo{author}{Arneodo, A.}, \bibinfo{author}{Bacry, E.},
  \bibinfo{author}{Graves, P.} \& \bibinfo{author}{Muzy, J.}
\newblock \bibinfo{title}{Characterizing long-range correlations in {DNA}
  sequencues from wavelet analysis}.
\newblock \emph{\bibinfo{journal}{Phys. Rev. Lett.}}
  \textbf{\bibinfo{volume}{74}}, \bibinfo{pages}{3293--3296}
  (\bibinfo{year}{1995}).

\bibitem{Machado2009}
\bibinfo{author}{Machado, R.} \& \bibinfo{author}{Weber, G.}
\newblock \bibinfo{title}{Wavelet coefficents as a guide to {DNA} phase
  transitions}.
\newblock \emph{\bibinfo{journal}{EPL}} \textbf{\bibinfo{volume}{87}},
  \bibinfo{pages}{38005} (\bibinfo{year}{2009}).

\bibitem{Grinsted2004}
\bibinfo{author}{Grinsted, A.}, \bibinfo{author}{Moore, J.} \&
  \bibinfo{author}{Jevrejeva, S.}
\newblock \bibinfo{title}{Application of the corss wavelet transform and
  wavelet cohorence to geophysical time series}.
\newblock \emph{\bibinfo{journal}{Nonlinear Process Geophys.}}
  \textbf{\bibinfo{volume}{11}}, \bibinfo{pages}{561--566}
  (\bibinfo{year}{2004}).

\bibitem{Hulata2000}
\bibinfo{author}{Hulata, E.}, \bibinfo{author}{Segev, R.},
  \bibinfo{author}{Shapira, Y.}, \bibinfo{author}{Benveniste, M.} \&
  \bibinfo{author}{Ben-Jacob, E.}
\newblock \bibinfo{title}{Detection and sorting of neural spikes using wavelet
  packets}.
\newblock \emph{\bibinfo{journal}{Phys. Rev. Lett.}}
  \textbf{\bibinfo{volume}{85}}, \bibinfo{pages}{4637--4640}
  (\bibinfo{year}{2000}).

\bibitem{Huang2003}
\bibinfo{author}{Huang, D.-W.}
\newblock \bibinfo{title}{Wavelet analysis of a traffic model}.
\newblock \emph{\bibinfo{journal}{Physica A}} \textbf{\bibinfo{volume}{329}},
  \bibinfo{pages}{298--308} (\bibinfo{year}{2003}).

\bibitem{Aguiar-Conraria2008}
\bibinfo{author}{Aguiar-Conraria, L.}, \bibinfo{author}{Azevedo, L.} \&
  \bibinfo{author}{Soares, M.}
\newblock \bibinfo{title}{Using wavelets to decompose the time-frequency
  effects of monetary policy}.
\newblock \emph{\bibinfo{journal}{Physica A}} \textbf{\bibinfo{volume}{387}},
  \bibinfo{pages}{2863--2878} (\bibinfo{year}{2008}).

\bibitem{Malkiel2003}
\bibinfo{author}{Malkiel, B.}
\newblock \bibinfo{title}{The efficient market hypothesis and its critics}.
\newblock \emph{\bibinfo{journal}{J. Econ. Perspect.}}
  \textbf{\bibinfo{volume}{17}}, \bibinfo{pages}{59-82}
  (\bibinfo{year}{2003}).

\bibitem{Rachev1999}
\bibinfo{author}{Rachev, S.} \& \bibinfo{author}{Weron, A.}
\newblock \bibinfo{title}{{CED} model for asset returns and fractal market
  hypothesis}.
\newblock \emph{\bibinfo{journal}{Math. Comput. Model.}}
  \textbf{\bibinfo{volume}{29}}, \bibinfo{pages}{23--36}
  (\bibinfo{year}{1999}).

\bibitem{Weron2000}
\bibinfo{author}{Weron, A.} \& \bibinfo{author}{Weron, R.}
\newblock \bibinfo{title}{Fractal market hypothesis and two power-laws}.
\newblock \emph{\bibinfo{journal}{Chaos Solitons Fractals}}
  \textbf{\bibinfo{volume}{11}}, \bibinfo{pages}{289--296}
  (\bibinfo{year}{2000}).

\bibitem{Farmer2004}
\bibinfo{author}{Farmer, J.}, \bibinfo{author}{Gillemot, L.},
  \bibinfo{author}{Lillo, F.}, \bibinfo{author}{Mike, S.} \&
  \bibinfo{author}{Sen, A.}
\newblock \bibinfo{title}{What really causes large price changes?}
\newblock \emph{\bibinfo{journal}{Quant. Financ.}}
  \textbf{\bibinfo{volume}{4}}, \bibinfo{pages}{383--397}
  (\bibinfo{year}{2004}).

\bibitem{Weber2006}
\bibinfo{author}{Weber, P.} \& \bibinfo{author}{Rosenow, B.}
\newblock \bibinfo{title}{Large stock price changes: volume or liquidity?}
\newblock \emph{\bibinfo{journal}{Quant. Financ.}}
  \textbf{\bibinfo{volume}{6}}, \bibinfo{pages}{7--14} (\bibinfo{year}{2006}).

\bibitem{Mike2008}
\bibinfo{author}{Mike, S.} \& \bibinfo{author}{Farmer, J.}
\newblock \bibinfo{title}{An empirical behavioral model of liquidity and
  volatility}.
\newblock \emph{\bibinfo{journal}{J. Econ. Dyn. Control}}
  \textbf{\bibinfo{volume}{32}}, \bibinfo{pages}{200--234}
  (\bibinfo{year}{2008}).

\bibitem{Roch2011}
\bibinfo{author}{Roch, A.}
\newblock \bibinfo{title}{Liquidity risk, price impacts and the replication
  problem}.
\newblock \emph{\bibinfo{journal}{Financ. Stoch.}}
  \textbf{\bibinfo{volume}{15}}, \bibinfo{pages}{399--419}
  (\bibinfo{year}{2011}).

\bibitem{Engle1982}
\bibinfo{author}{Engle, R.}
\newblock \bibinfo{title}{Autoregressive conditional heteroskedasticity with
  estimates of variance of {United Kingdon} inflation}.
\newblock \emph{\bibinfo{journal}{Econometrica}} \textbf{\bibinfo{volume}{50}},
  \bibinfo{pages}{987--1008} (\bibinfo{year}{1982}).

\bibitem{Bollerslev1986}
\bibinfo{author}{Bollerslev, T.}
\newblock \bibinfo{title}{Generalized autoregressive conditional
  heteroskedasticity}.
\newblock \emph{\bibinfo{journal}{J. Econom.}}
  \textbf{\bibinfo{volume}{31}}, \bibinfo{pages}{307--327}
  (\bibinfo{year}{1986}).

\bibitem{Kristoufek2012}
\bibinfo{author}{Kristoufek, L.}
\newblock \bibinfo{title}{Fractal markets hypothesis and the global financial
  crisis: Scaling, investment horizons and liquidity}.
\newblock \emph{\bibinfo{journal}{Adv. Complex Syst.}}
  \textbf{\bibinfo{volume}{15}}, \bibinfo{pages}{1250065}
  (\bibinfo{year}{2012}).

\bibitem{Allen1996}
\bibinfo{author}{Allen, M.} \& \bibinfo{author}{Smith, L.}
\newblock \bibinfo{title}{{Monte Carlo SSA}: detecting irregular oscillations
  in the presence of coloured noise}.
\newblock \emph{\bibinfo{journal}{J. Clim.}}
  \textbf{\bibinfo{volume}{9}}, \bibinfo{pages}{3373--3404}
  (\bibinfo{year}{1996}).

\bibitem{Percival2000}
\bibinfo{author}{Percival, D.} \& \bibinfo{author}{Walden, A.}
\newblock \emph{\bibinfo{title}{Wavelet Methods for Time series Analysis}}
  (\bibinfo{publisher}{Cambridge University Press}, \bibinfo{year}{2000}).

\bibitem{Rua2009}
\bibinfo{author}{Rua, A.} \& \bibinfo{author}{Nunes, L.}
\newblock \bibinfo{title}{International comovement of stock market returns: A
  wavelet analysis}.
\newblock \emph{\bibinfo{journal}{J. Empir. Financ.}}
  \textbf{\bibinfo{volume}{16}}, \bibinfo{pages}{632--639}
  (\bibinfo{year}{2009}).

\bibitem{Vacha2012}
\bibinfo{author}{Vacha, L.} \& \bibinfo{author}{Barunik, J.}
\newblock \bibinfo{title}{Comovement of energy commodities revisited: evidence
  from wavelet coherence analysis}.
\newblock \emph{\bibinfo{journal}{Energy Econ.}}
  \textbf{\bibinfo{volume}{34}}, \bibinfo{pages}{241--247}
  (\bibinfo{year}{2012}).

\bibitem{Vacha2013}
\bibinfo{author}{Vacha, L.}, \bibinfo{author}{Janda, K.},
  \bibinfo{author}{Kristoufek, L.} \& \bibinfo{author}{Zilberman, D.}
\newblock \bibinfo{title}{Time-frequency dynamics of biofuel-fuel-food system}.
\newblock \emph{\bibinfo{journal}{Energy Econ.}}
  \textbf{\bibinfo{volume}{40}}, \bibinfo{pages}{233--241}
  (\bibinfo{year}{2013}).

\end{thebibliography}

\section*{Acknowledgements}
The support from the Grant Agency of the Czech Republic (GACR) under projects P402/11/0948 and 402/09/0965, Grant Agency of Charles University (GAUK) under project $1110213$ are gratefully acknowledged. We also thank Grinsted, Moore \& Jevrejeva \cite{Grinsted2004} for the MatLab package for continuous wavelet analysis.

\section*{Author contributions}
L.K. solely wrote the main manuscript text, prepared the figures and reviewed the manuscript.

\section*{Additional information}
\textbf{Competing financial interests:} The author declares no competing financial interests.\\
\textbf{Data retrieval:} Data were obtained from http://finance.yahoo.com on 8.6.2013. \\

\end{document}